\documentclass[12pt]{article}
\usepackage{amssymb}
\begin{document}
\topmargin-.4cm 
\begin{titlepage}
\begin{flushright}
ETH-TH/97-16
\end{flushright}
\vskip.3in
\begin{center}{\Large\bf
An Effective  Superstring Spectral Action }   
\end{center}
\vskip.3in
\begin{center}{
 Ali H. Chamseddine\\
 Theoretische Physik\\
 ETH-Z\"urich, CH-8093 } 
\end{center}
\vskip.5cm
\begin{center}
{\bf Abstract}
\end{center}
\begin{quote}
A supersymmetric theory in two-dimensions has enough data to 
define a noncommutative space thus making it possible to use
all tools of noncommutative geometry. In particular, we
apply this to the $N=1$ supersymmetric non-linear sigma model 
and derive an expression for the generalized loop space 
Dirac operator, in presence of
a general background, using canonical quantization.  
The spectral action  principle is used to show that the superstring partition function is also
a spectral action valid for the fluctuations of the string modes.
\end{quote}
\end{titlepage}
\newpage 

\section{\protect\medskip Introduction}

One of the most important results in string theory is the appearance of the
graviton in the spectrum of the closed string. The nature of the
gravitational force is then deduced by studying the graviton-graviton
scattering in the S-matrix using the graviton vertex operator in conformal
field theory. Similarly, the interaction of the other string spectra,
massless or massive, could be obtained. Alternatively, if the propagation of
the string is determined from the variation of a two-dimensional non-linear
sigma model with curved background geometry the renormalizability of the
theory dictates that the background metric satisfies generalized Einstein
equations \cite{Luis81}. Some attempts were made to understand string theory
based on loop space geometry \cite{BR}, but this had limitations because the
needed geometrical tools are not available.

There is, however, a close relation between two-dimensional systems defining
what is known as supersymmetric quantum mechanics \cite{W82} \cite{W85} and
noncommutative geometry \cite{ConnesBook}. A supersymmetric theory in two
dimensions are classified by the number of left and right supersymmetries
denoted by $(p,q)$. Invariance under each of these supersymmetries gives a
conserved charge. The most familiar examples are $(1,1)$ supersymmetry of
the superstring and $(1,0)$ supersymmetry of the heterotic string. The
supersymmetry charges are considered as differential operators on the loop
space $\Omega (M)$ \cite{W82}. To make the connection with noncommutative
geometry one has to identify the spectral triple $({\mathcal{A}},\mathcal{H}%
,D)\ $where ${\mathcal{A}}$ is an algebra of operators, $\mathcal{H}$ a
Hilbert space and $D$ a Dirac operator acting on $\mathcal{H}.$ The presence
of more than one supersymmetry charge implies restrictions on the geometry.
The algebra $\mathcal{A}$ can be identified with the superconformal algebra
associated with a two-dimensional model and $\mathcal{H}$ with the Hilbert
space of states \cite{CF} \cite{FG}.

Within the formalism of noncommutative geometry, it is possible to define,
in complete analogy with Riemannian geometry, the metric, distance,
connection, torsion, etc. \cite{CFFgravity}. Because of the infinite
dimensional nature of such spaces many ambiguities are expected. Describing
the dynamics of such a system geometrically is not a trivial exercise.

With every supersymmetric system we can associate a triple $(\mathcal{A},%
\mathcal{H},D)$ from which one can define a noncommutative space. The
dynamics of the spectrum of these theories is system dependent. The
two-dimensional system that we will consider for our study is the
supersymmetric non-linear sigma model coupled to background fields \cite
{Luis81}. We shall evaluate the conserved supersymmetric currents and the
associated charges $Q_{+}$ and $Q_{-}$. We first identify the algebra and
Hilbert space and then study some properties of this space. From the fact
that the spectral properties of $(\mathcal{A},\mathcal{H},Q_{\pm })$ encodes
the necessary geometric properties, we shall postulate a spectral action
that will give the correct dynamics for all the superstring spectrum \cite
{ACAC}. The results of this work were given without any details in reference 
\cite{Ali}.

The plan of this paper is as follows. In the second section we review the
supersymmetric non-linear sigma model and derive the full expression of the
supersymmetric charges in the presence of torsion. We also evaluate the
Hamiltonian and momentum after the system is quantized. In section three we
give a review of noncommutative geometry. In section four we use these tools
to extract information about the geometry of the superstring noncommutative
space. We also postulate a spectral action that describes the dynamics of
the superstring spectrum. We show that this action gives, in the low-energy
limit, the superstring effective action. Section five is the conclusion.

\section{Supersymmetric quantum mechanics and the non-linear sigma model}

A supersymmetric theory in two-dimensions have supersymmetry generators
associated with it. These transform bosonic states into fermionic states.
What is special about two-dimensions is that one can split right-movers from
left-movers and therefore can have an asymmetry between the supersymmetry in
both sectors. The number of supersymmetries is denoted by $(p,q)$. The
supersymmetry generators satisfy the algebra 
\begin{eqnarray}
\left\{ Q_{+i},Q_{+j}\right\} &=&\delta _{ij}P_{z}  \nonumber \\
\left\{ Q_{-i},Q_{-j}\right\} &=&\delta _{ij}P_{\overline{z}} \\
\left[ J,Q_{\pm i}\right] &=&\mp Q_{\pm i}  \nonumber
\end{eqnarray}

In this paper we shall only consider the simple case of $(1,1)$
supersymmetry. This is easily realized by starting with the supersymmetric
non-linear sigma model where the matter superfields are $\Phi ^{\mu }(\xi
,\theta _{+},\theta _{-})$ where $\xi ^{\alpha }$ are the two-dimensional
space-time coordinates, $\xi ^{0}=\tau $ and $\xi ^{1}=\sigma .$ These
fields interact non-linearly with backgrounds of the form $G_{\mu \nu
}\left[ \Phi \right] $, $B_{\mu \nu }\left[ \Phi \right] $, $\varphi \left[
\Phi \right] $ where $G_{\mu \nu }$ is a symmetric tensor $B_{\mu \nu }$ is
antisymmetric and $\varphi $ is a scalar. More general interactions could be
introduced. In what follows and for simplicity, we shall set the dilaton
background to zero. Using superconformal invariance one can choose the
superconformal gauge which fixes the superzweibein, $E_{M}^{A}$ to be: 
\begin{equation}
\left( 
\begin{array}{ll}
E_{\alpha }^{a}=\delta _{\alpha }^{a} & E_{\alpha }^{m}=0 \\ 
E_{r}^{a}=\left( \gamma ^{a}\theta \right) _{r} & E_{r}^{m}=\delta _{\alpha
}^{m}
\end{array}
\right)
\end{equation}
where $r=+,-.$ In this gauge the superspace action is given by 
\begin{equation}
S=\frac{T}{2}\int d^{2}\xi d\theta _{+}d\theta _{-}\left( (G_{\mu \nu
}\left[ \Phi \right] +B_{\mu \nu }\left[ \Phi \right] \ )D_{+}\Phi ^{\mu
}D_{-}\Phi ^{\nu }\right) \   \label{suaction}
\end{equation}
where $D_{\pm }=\frac{\partial }{\partial \theta _{\pm }}-i\theta _{\pm
}\partial _{\pm }$ and $\partial _{\pm }=\partial _{0}\pm \partial _{1}$.
These satisfy $D_{\pm }^{2}=-i\partial _{\pm }$.

The superfield expansion of $\Phi ^{\mu }(\xi ,\theta _{+},\theta _{-})$ is 
\begin{equation}
\Phi ^{\mu }(\xi ,\theta _{+},\theta _{-})=X^{\mu }(\xi )+i\theta _{+}\psi
^{\mu +}(\xi )-i\theta _{-}\psi ^{\mu -}(\xi )+i\theta _{+}\theta _{-}F^{\mu
}(\xi )
\end{equation}
Expanding $G_{\mu \nu }\left[ \Phi \right] $ ,$B_{\mu \nu }\left[ \Phi
\right] $ in terms of components one finds that the action $S$ becomes a
function of $G_{\mu \nu }\left[ X\right] ,$ $B_{\mu \nu }\left[ X\right] $
and their derivatives as well as $\psi ^{\mu \pm }$ and $F^{\mu }$. This
gives, after integrating over $\theta _{\pm }$: 
\begin{equation}
\begin{array}{c}
S=\frac{T}{2}\int d^{2}\xi \left( (G_{\mu \nu }+B_{\mu \nu })(-i\partial
_{-}\psi ^{\mu +}\psi ^{\nu +}+i\psi ^{\mu -}\partial _{+}\psi ^{\nu
-}+\partial _{-}X^{\mu }\partial _{+}X^{\nu }+F^{\mu }F^{\nu })\right. \quad 
\\ 
\quad \quad \quad \qquad \qquad \!-i(G_{\mu \nu ,\rho }+B_{\mu \nu ,\rho
})\psi ^{\rho +}(\partial _{-}X^{\mu }\psi ^{\nu +}-\psi ^{\mu -}F^{\upsilon
})\qquad \qquad \qquad \qquad \qquad \qquad  \\ 
\qquad \qquad \quad -i(G_{\mu \nu ,\rho }+B_{\mu \nu ,\rho })\psi ^{\rho
-}(\psi ^{\mu -}\partial _{+}X^{\nu }+F^{\mu }\psi ^{\nu +})\qquad \qquad
\qquad \qquad \qquad  \\ 
\qquad \qquad \qquad \qquad +\left. \psi ^{\mu -}\psi ^{\nu +}((G_{\mu \nu
,\rho }+B_{\mu \nu ,\rho })F^{\rho }+(G_{\mu \nu ,\rho \sigma }+B_{\mu \nu
,\rho \sigma })\psi ^{\rho -}\psi ^{\sigma +})\right) \quad \quad \quad
\qquad 
\end{array}
\label{action}
\end{equation}
Eliminating the auxiliary fields $F^{\mu }$ by its equations of motion
gives: 
\begin{equation}
F^{\rho }=\pm i\psi ^{\mu -}\psi ^{\nu +}\left( \Gamma _{\mu \nu }^{\quad
\rho }-\frac{1}{2}H_{\mu \nu }^{\quad \rho }\right)   \label{aux}
\end{equation}
where $\Gamma _{\mu \nu }^{\rho \quad }$ is the Christoffel symbol: 
\begin{equation}
\Gamma _{\mu \nu }^{\quad \rho \quad }=\frac{1}{2}G^{\rho \kappa }(G_{\mu
\kappa ,\nu }+G_{\nu \kappa ,\rho }-G_{\mu \nu ,\kappa })
\end{equation}
and $H_{\mu \nu \rho }$ is the field strength of $B_{\mu \nu }$ : 
\begin{equation}
H_{\mu \nu \rho }=(B_{\mu \nu ,\rho }+B_{v\rho ,\mu }+B_{\rho \mu ,\nu })
\end{equation}
Since $F^{\mu }$ appears quadratically, performing the gaussian integration
is equivalent to substituting the value of $F^{\mu }$ given in equation (\ref
{aux}) into the action (\ref{action}). It will prove convenient to define
fermi fields $\psi ^{\pm a}$ with tangent space-indices: $\psi ^{\pm
a}=e_{\mu }^{a}[X]\psi ^{\pm \mu }$ where $e_{\mu }^{a}[X]$ is the square
root of $G_{\mu \nu }[X]$: 
\begin{equation}
G_{\mu \nu }[X]=e_{\mu }^{a}[X]\,\eta _{ab}\,e_{\nu }^{b}[X]
\end{equation}
and $\eta _{ab}$ is the flat space metric (Minkowski or Euclidean ). After
some manipulations and integrating by parts we get \cite{Luis81} 
\begin{eqnarray}
S &=&\frac{T}{2}\int d^{2}\xi \,\left( \left( G_{\mu \nu }[X]+B_{\mu \nu
}[X]\right) \partial _{-}X^{\mu }\partial _{+}X^{\nu }\right.   \nonumber \\
&&+i\psi ^{a+}\left( \eta _{ab}\partial _{-}+\omega _{\mu ab}^{+}\partial
_{-}X^{\mu }\right) \psi ^{b+}  \nonumber \\
&&+i\psi ^{a-}\left( \eta _{ab}\partial _{+}+\omega _{\mu ab}^{-}\partial
_{+}X^{\mu }\right) \psi ^{b-}  \nonumber \\
&&+\frac{1}{2}\psi ^{a+}\psi ^{b+}\psi ^{c-}\psi ^{d-}R_{cdab}^{+}[X] 
\nonumber \\
&&\left. -\frac{i}{2}\partial _{-}\left( B_{\mu \nu }[X]\psi ^{\mu +}\psi
^{\nu +}\right) +\frac{i}{2}\partial _{+}\left( B_{\mu \nu }[X]\psi ^{\mu
-}\psi ^{\nu -}\right) \right)   \label{surface}
\end{eqnarray}
where $\omega _{\mu }^{\pm ab}=\omega _{\mu }^{\,\,ab}\pm \frac{1}{2}H_{\mu
}^{\,\,ab}$ and $\omega _{\mu }^{\,\,ab}$ is related to $e_{\mu }^{a}$
through the torsion free condition 
\begin{equation}
\partial _{\mu }e_{\nu }^{a}+\omega _{\mu }^{\,\,ab}e_{\nu b}-\Gamma _{\mu
\nu }^{\quad \rho }e_{\rho }^{a}=0
\end{equation}
which can be completely solved for $\omega _{\mu }^{\,\,ab}$. The Riemannian
tensors $R_{\ \nu \rho \sigma }^{\pm \mu }$ (with torsion) are defined by: 
\begin{equation}
R_{\ \quad \nu \rho \sigma }^{\pm \mu }=\partial _{\rho }\Gamma _{\sigma \nu
}^{\pm \,\,\mu }+\Gamma _{\rho \kappa }^{\pm \,\mu }\Gamma _{\sigma \nu
}^{\pm \,\kappa }-\left( \rho \leftrightarrow \sigma \right) 
\end{equation}
and satisfy 
\begin{equation}
R_{\,\mu \upsilon \rho \sigma }^{+}=-R_{\nu \mu \rho \sigma }^{+}=-R_{\,\mu
\upsilon \sigma \rho }^{+}=R_{\,\rho \sigma \mu \nu }^{-}
\end{equation}
curved indices are changed to flat ones with the help of $e_{\mu }^{a}$ and
its inverse. We have kept total derivative terms in the action as these will
be important in evaluating the supercharges.

To determine the supercurrents we first write the supersymmetry
transformation of the superfield $\Phi ^{\mu }$ : 
\begin{equation}
\delta \Phi ^{\mu }=(\epsilon _{+}D_{+}+\epsilon _{-}D_{-})\Phi ^{\mu }
\end{equation}
which gives for the components

\begin{eqnarray}
\!\delta X^{\mu } &=&i(\epsilon _{+}\psi ^{\mu +}-\epsilon _{-}\psi ^{\mu
-})\quad \qquad  \nonumber \\
\delta \psi ^{\mu +} &=&-\epsilon _{+}\partial _{+}X^{\mu }+\epsilon
_{-}F^{\mu }\qquad \quad  \nonumber \\
\delta \psi ^{\mu -} &=&\epsilon _{-}\partial _{-}X^{\mu }+\epsilon
_{+}F^{\mu }  \nonumber \\
\delta F^{\mu } &=&-i(\epsilon _{+}\partial _{+}\psi ^{\mu -}+\epsilon
_{-}\partial _{-}\psi ^{\mu +})\qquad \qquad  \label{trans}
\end{eqnarray}
Noting that the supersymmetric variation of the Lagrangian in (\ref{surface}%
) is a total derivative, we have 
\begin{equation}
\begin{array}{c}
\frac{1}{T}\delta \mathcal{L}=\frac{i}{2}\epsilon _{+}\partial _{+}\left(
(G_{\mu \nu }+B_{\mu \nu })\left( \partial _{-}X^{\mu }\psi ^{\nu +}-\psi
^{\mu +}F^{\nu }\right) \right. \\ 
\quad \quad \quad \quad \quad \left. -(iG_{\mu \nu ,\rho }+B_{\mu \nu ,\rho
})\psi ^{\rho -}\psi ^{\mu -}\psi ^{\nu +})\right) \\ 
\qquad \qquad 
\begin{array}{c}
-\frac{i}{2}\epsilon _{-}\partial _{-}\left( (G_{\mu \nu }+B_{\mu \nu
})\left( \psi ^{\mu -}\partial _{+}X^{\nu }+F^{\mu }\psi ^{\nu +}\right)
\right. \\ 
\quad \left. -i(G_{\mu \nu ,\rho }+B_{\mu \nu ,\rho })\psi ^{\rho +}\psi
^{\mu +}\psi ^{\nu -})\right)
\end{array}
\\ 
\equiv \epsilon _{+}\partial _{+}\kappa ^{+}+\epsilon _{-}\partial
_{-}\kappa ^{-}\qquad \qquad \qquad \quad \,
\end{array}
\label{var}
\end{equation}
The supercharges $j_{\pm }$ are then defined by 
\begin{equation}
\frac{1}{T}(\pm ij_{\pm })=\frac{\delta X^{\mu }}{\delta \epsilon _{\pm }}%
\frac{\delta \mathcal{L}}{\delta \partial _{0}X^{\mu }}+\frac{\delta \psi
^{\mu +}}{\delta \epsilon _{\pm }}\frac{\delta \mathcal{L}}{\delta \partial
_{0}\psi ^{\mu +}}+\frac{\delta \psi ^{\mu -}}{\delta \epsilon _{\pm }}\frac{%
\delta \mathcal{L}}{\delta \partial _{0}\psi ^{\mu -}}+\kappa ^{\pm }.
\label{jpm}
\end{equation}
A simple calculation gives 
\begin{equation}
\frac{i}{T}(\pm j_{\pm })=\pm i\psi ^{\mu \pm }G_{\mu \nu }\partial _{\pm
}X^{\nu }-\frac{1}{6}\psi ^{\mu \pm }\psi ^{\nu \pm }\psi ^{\rho \pm }H_{\mu
\nu \rho }.  \label{cur}
\end{equation}
The non-linear sigma model must now be quantized \cite{W82} \cite{Luis83}.
As the fermions form a first order system, they will be constrained, and
Poisson brackets has to be replaced with Dirac brackets. Denoting 
\begin{equation}
\tau ^{a\pm }\equiv \frac{\delta \mathcal{L}}{\delta \partial _{0}\psi
^{a\pm }}=-\frac{i}{2}T\psi ^{a\pm },
\end{equation}
the constraint equations are 
\begin{equation}
\chi ^{a\pm }=\tau ^{a\pm }+\frac{i}{2}T\psi ^{a\pm }=0
\end{equation}
From the Poisson bracket 
\begin{equation}
\left\{ \psi ^{a\pm },\tau _{b}^{\pm }\right\} _{P.B}=-\delta _{a}^{b}
\end{equation}
one can verify that 
\begin{equation}
\left\{ \chi _{a}^{\pm },\chi _{b}^{\pm }\right\} =-iT\delta _{ab}\equiv
C_{ab}
\end{equation}
and it is inconsistent to impose the constraint. This is remedied by
introducing the Dirac bracket \cite{ Davies} 
\begin{equation}
\left\{ A,B\right\} _{D.B}=\left\{ A,B\right\} _{P.B}-\left\{ A,\chi
_{a}\right\} _{P.B}C_{ab}^{-1}\left\{ B,\chi _{b}\right\} _{P.B}
\end{equation}
This now gives: 
\begin{equation}
\left\{ \psi ^{a\pm },\psi ^{b\pm }\right\} _{DB}=-\frac{i}{T}\delta ^{ab}.
\end{equation}
Quantization is carried in replacing Poisson and Dirac brackets by
equal-time commutators and anticommutators: 
\begin{eqnarray}
\left[ X^{\mu }(\sigma ,\tau ),P_{\nu }(\sigma ^{^{\prime }},\tau )\right]
&=&i\delta _{\mu }^{\nu }\delta (\sigma -\sigma ^{\prime }) \\
\left\{ \psi ^{a\pm }(\sigma ,\tau ),\psi ^{b\pm }(\sigma ^{\prime },\tau
)\right\} &=&\frac{2}{T}\eta ^{ab}\delta (\sigma -\sigma ^{\prime })
\end{eqnarray}
Rotating the fermions to the chiral basis: 
\begin{eqnarray}
\psi ^{a+} &=&\frac{1}{\sqrt{T}}(\chi ^{a}+\overline{\chi }^{a}) \\
\psi ^{a-} &=&\frac{i}{\sqrt{T}}(\chi ^{a}-\overline{\chi }^{a})
\end{eqnarray}
implies that 
\begin{equation}
\left\{ \chi ^{a}(\sigma ,\tau ),\overline{\chi }^{b}(\sigma ^{\prime },\tau
)\right\} =\frac{1}{2}\eta ^{ab}\delta (\sigma -\sigma ^{\prime })
\label{clif}
\end{equation}
The momentum $P_{\nu }(\sigma ,\tau )$ can be realized by acting on the
space $X^{\mu }(\sigma ,\tau )$ through the relation 
\begin{equation}
P_{\mu }(\sigma ,0)=-i\frac{\delta }{\delta X^{\mu }(\sigma ,0)}  \label{mom}
\end{equation}
where the time $\tau $ is set to zero. Using the definition of $P_{\mu }=%
\frac{\delta \mathcal{L}}{\delta \partial _{0}X^{\mu }}$ one finds 
\begin{equation}
P_{\mu }=T\left( G_{\mu \nu }\partial _{0}X^{\nu }+\frac{i}{2}\psi ^{a+}\psi
^{b+}\omega _{\mu ab}^{+}+\frac{i}{2}\psi ^{a-}\psi ^{b-}\omega _{\mu
ab}^{-}\right)  \label{peq}
\end{equation}
Substituting equation (\ref{peq}) for $P_{\mu }$ into equation (\ref{jpm})
for $j_{\pm }$, the currents then take the simple form 
\begin{eqnarray}
j_{+}\left( \sigma \right) &=&-\frac{i}{\sqrt{T}}\chi ^{a}(\sigma )\left(
e_{a}^{\mu }[X]\nabla _{\mu }+\frac{1}{3}H_{abc}\chi ^{b}(\sigma )\chi
^{c}(\sigma )\right)  \nonumber \\
&&+\sqrt{T}\left( \overline{\chi }^{a}(\sigma )e_{\nu a}[X]-\chi ^{a}(\sigma
)e_{a}^{\mu }[X]B_{\mu \nu }[X]\right) \frac{dX^{\nu }}{d\sigma }
\label{jplus} \\
j_{-}\left( \sigma \right) &=&-\frac{i}{\sqrt{T}}\overline{\chi }^{a}(\sigma
)\left( e_{a}^{\mu }[X]\nabla _{\mu }+\frac{1}{3}H_{abc}\overline{\chi }%
^{b}(\sigma )\overline{\chi }^{c}(\sigma )\right)  \nonumber \\
&&+\sqrt{T}\left( \chi ^{a}(\sigma )e_{\nu a}[X]-\overline{\chi }^{a}(\sigma
)e_{a}^{\mu }[X]B_{\mu \nu }[X]\right) \frac{dX^{\nu }}{d\sigma }\quad
\label{jminus}
\end{eqnarray}
where we have rotated the currents $j_{\pm }$ to the chiral basis 
\begin{eqnarray}
j &=&\frac{1}{2}(j_{+}-ij_{-}) \\
\overline{j} &=&\frac{1}{2}(j_{+}+ij_{-})
\end{eqnarray}
The covariant derivative $\nabla _{\mu }$ is defined by 
\begin{equation}
\nabla _{\mu }=\frac{\delta }{\delta X^{\mu }}+\omega _{\mu ab}[X]\left(
\chi ^{a}(\sigma )\overline{\chi }^{b}(\sigma )+\overline{\chi }^{a}(\sigma
)\chi ^{b}(\sigma )\right)  \label{cov}
\end{equation}
and one must normal order to avoid the ambiguity of multiplying fields at
the same point. The conserved supersymmetry charges are 
\begin{eqnarray}
Q &=&\int\limits_{0}^{2\pi }d\sigma \,j(\sigma ) \\
\overline{Q} &=&\int\limits_{0}^{2\pi }d\sigma \,\overline{j}(\sigma )
\end{eqnarray}
It is a tedious exercise to show that after quantization, the charges $Q$
and $\overline{Q}$ form a supersymmetry algebra with the properties 
\begin{equation}
Q^{2}=\frac{1}{2}P=\overline{Q}^{2}  \label{momentum}
\end{equation}
\begin{equation}
\left\{ Q,\overline{Q}\right\} =\frac{1}{2}H  \label{ham}
\end{equation}
where $P$ is the momentum generating reparametrizations on the circle and $H$
is the Hamiltonian.

We shall follow the strategy adopted in references \cite{Davies} \cite
{Braden} in the grouping of the terms where relations (\ref{momentum},\ref
{ham}) were proved for the case of a point particle. The important point to
realize here is that the fields $e_{\mu }^{a}$ and $B_{\mu \nu }$ have
functional dependence on $X^{\mu }$. The relation 
\begin{equation}
\frac{\delta }{\delta X^{\mu }(\sigma )}f\,\,\left[ X(\sigma )\right] =\frac{%
\delta f}{\delta X^{\mu }}\delta (\sigma -\sigma ^{\prime })
\end{equation}
is used frequently. The two-dimensional momentum $P$ is given by 
\begin{equation}
P=-i\,\int\limits_{0}^{2\pi }d\sigma \frac{dX^{\mu }}{d\sigma }\nabla _{\mu
}+2i\int\limits_{0}^{2\pi }d\sigma \chi _{a}(\sigma )\frac{D\overline{\chi }%
^{a}}{D\sigma }  \label{Peq}
\end{equation}
and the covariant derivative $\frac{D}{D\sigma }$ is defined by 
\begin{equation}
\frac{D\overline{\chi }^{a}}{D\sigma }=\frac{d\overline{\chi }^{a}}{d\sigma }%
+\frac{dX^{\mu }}{d\sigma }\omega _{\mu \ b}^{\ a\ }[X]\overline{\chi }%
^{b}(\sigma )
\end{equation}

Note that in the limit when $X^{\mu }(\sigma )$ and $\chi ^{a}(\sigma )$
become independent of $\sigma $, $P$ as given in equation (\ref{Peq}) will
vanish automatically. A very lengthy calculation gives for the Hamiltonian: 
\begin{eqnarray}
H &=&-\frac{1}{2T}\int\limits_{0}^{2\pi }d\sigma \left[ \left( \nabla
^{a}\nabla _{a}+\omega _{b}^{\ ab}[X(\sigma )]\nabla _{a}+4\chi ^{a}(\sigma )%
\overline{\chi }^{b}(\sigma )\chi ^{c}(\sigma )\overline{\chi }^{d}(\sigma
)R_{abcd}[X(\sigma )]\right) \right.  \nonumber \\
&&\;\;\qquad \quad +\frac{2}{3}\left( \chi ^{a}(\sigma )\overline{\chi }%
^{b}(\sigma )\overline{\chi }^{c}(\sigma )\overline{\chi }^{d}(\sigma )+%
\overline{\chi }^{a}(\sigma )\chi ^{b}(\sigma )\chi ^{c}(\sigma )\chi
^{d}(\sigma )\right) \nabla _{a}H_{bcd}[X(\sigma )]  \nonumber \\
&&\qquad \qquad +\left( \chi ^{b}(\sigma )\chi ^{c}(\sigma )+\overline{\chi }%
^{b}(\sigma )\overline{\chi }^{c}(\sigma )\right) H_{\ bc}^{a}[X(\sigma
)]\nabla _{a}  \nonumber \\
&&\qquad \qquad \quad +\frac{1}{3}\left( \chi ^{a}(\sigma )\chi ^{b}(\sigma )%
\overline{\chi }^{c}(\sigma )\overline{\chi }^{d}(\sigma )+\chi ^{a}(\sigma )%
\overline{\chi }^{b}(\sigma )\overline{\chi }^{c}(\sigma )\chi ^{d}(\sigma
)\right.  \nonumber \\
&&\qquad \qquad \quad +\left. \overline{\chi }^{a}(\sigma )\overline{\chi }%
^{b}(\sigma )\chi ^{c}(\sigma )\chi ^{d}(\sigma )H_{ab}^{\ \ e}[X(\sigma
)]H_{ecd}[X(\sigma )]\right)  \nonumber \\
&&\qquad \qquad -2iT\left( \chi _{a}\frac{D{\chi }^{a}}{D\sigma }+\overline{%
\chi }_{a}\frac{D\overline{\chi }^{a}}{D\sigma }\right)  \nonumber \\
&&\qquad \qquad -2iT\left( \chi ^{a}\frac{D\chi ^{b}}{D\sigma }+\overline{%
\chi }^{a}\frac{D\overline{\chi }^{b}}{D\sigma }\right) B_{ab}[X(\sigma )] 
\nonumber \\
&&-2iTe_{a}^{\mu }[X(\sigma )]e_{b}^{\nu }[X(\sigma )]\left( \chi
^{b}(\sigma )\overline{\chi }^{a}(\sigma )+\overline{\chi }^{b}(\sigma )\chi
^{a}(\sigma )\right) \left( \nabla _{\rho }B_{\mu \nu }-\nabla _{\nu }B_{\mu
\rho }\right) \frac{dX^{\rho }}{d\sigma }  \nonumber \\
&&+2iTB_{\mu \nu }\frac{dX^{\nu }}{d\sigma }\nabla ^{\mu }-2i\left( \chi
^{a}(\sigma )\chi ^{b}(\sigma )+\overline{\chi }^{a}(\sigma )\overline{\chi }%
^{b}(\sigma )\right) H_{ab}^{\ \ c}[X(\sigma )]B_{c\nu }[X(\sigma )]\frac{%
dX^{\nu }}{d\sigma }  \nonumber \\
&&\qquad \qquad \left. -T^{2}\left( G_{\mu \nu }[X(\sigma )]+B_{\mu \kappa
}[X(\sigma )]B_{\ \nu }^{\kappa }[X(\sigma )]\right) \frac{dX^{\mu }}{%
d\sigma }\frac{dX^{\nu }}{d\sigma }\right] \!\!\!\!\qquad
\label{Hamiltonian}
\end{eqnarray}
Again, In the limit when $X^{\mu }(\sigma )$ and $\chi ^{a}(\sigma )$ become
independent of $\sigma $, the expression of the Hamiltonian in equation (\ref
{Hamiltonian}) reduces to that derived in \cite{Braden}

The time development of the coordinates $X^{\mu }(\sigma ,\tau )$ is
governed by the equation 
\begin{equation}
X^{\mu }(\sigma ,\tau )=e^{-\tau H}X^{\mu }(\sigma ,0)e^{\tau H}
\end{equation}
which for a general background is very complicated. Assuming the boundary
conditions that $X^{\mu }(\sigma ,0)$ are periodic for the closed string,
one gets 
\begin{equation}
X^{\mu }(\sigma )=X_{0}^{\mu }+\sum_{n>0}\frac{1}{\sqrt{n\pi T}}\left(
a_{n}^{\mu }\cos n\sigma +\widetilde{a}_{n}^{\mu }\sin n\sigma \right)
\label{Xosc}
\end{equation}
The Hamiltonian is a function of an even number of $\chi ^{a}(\sigma )$ and $%
\overline{\chi }^{a}(\sigma )$, therefore one can have either periodic or
antiperiodic boundary conditions for the fermions, giving rise to a Ramond
sector (R) and a Neveu-Schwarz (NS) sector respectively. This allows for the
functions $\chi ^{a}(\sigma )$ and $\overline{\chi }^{a}(\sigma )$ to be
expanded in the form: 
\begin{eqnarray}
\chi ^{a}(\sigma ) &=&\frac{1}{\sqrt{2\pi }}\sum_{r\in Z_{0}+\phi }\left(
c_{r}\cos r\sigma +d_{r}\sin r\sigma \right)  \label{chi} \\
\overline{\chi }^{a}(\sigma ) &=&\frac{1}{\sqrt{2\pi }}\sum_{r\in Z_{0}+\phi
}\left( \overline{c}_{r}\cos r\sigma +\overline{d}_{r}\sin r\sigma \right)
\label{chibar}
\end{eqnarray}
where $\phi =0$ for the R-sector and $\phi =\frac{1}{2}$ for the NS-sector.
The momentum $P_{\mu }=-i\frac{\delta }{\delta X^{\mu }(\sigma )}$ can also
be expanded in terms of oscillators: 
\begin{equation}
P_{\mu }=-i\left( \frac{1}{2\pi }\frac{\delta }{\delta X_{0}^{\mu }}%
+\sum_{n>0}\sqrt{\frac{nT}{\pi }}\left( \frac{\delta }{\delta a_{n}^{\mu }}%
\cos n\sigma +\frac{\delta }{\delta \widetilde{a}_{n}^{\mu }}\sin n\sigma
\right) \right)
\end{equation}
The quantization conditions on the fermions imply that the only
non-vanishing anti-commutators are 
\begin{equation}
\begin{array}{ccc}
\left\{ c_{r}^{a},\overline{c}_{s}^{b}\right\} =2\delta _{rs}\eta ^{ab} & 
r,s\neq 0 &  \\ 
\left\{ d_{r}^{a},\overline{d}_{s}^{b}\right\} =2\delta _{rs}\eta ^{ab} &  & 
\end{array}
\label{cdos}
\end{equation}
while the fermionic zero modes occur only in the R-sector and satisfy the
anticommutation relations: 
\begin{equation}
\left\{ c_{0}^{a},\overline{c}_{0}^{b}\right\} =\eta ^{ab}  \label{c0cliff}
\end{equation}
Therefore both $c_{0}^{a}+\overline{c}_{0}^{a}$ and $i(c_{0}^{a}-\overline{c}%
_{0}^{a})$ generate Clifford algebras, and give rise to creation and
annihilation operators for the vacuum state.

At this point it is useful to make contact with the case when the background
geometry is flat: 
\begin{equation}
G_{\mu \nu }=\eta _{\mu \nu },\qquad B_{\mu \nu }=0
\end{equation}
In this case the Hamiltonian simplifies to: 
\begin{equation}
H=-\frac{1}{2T}\int\limits_{0}^{2\pi }d\sigma \left[ \frac{\delta }{\delta
X^{\mu }}\frac{\delta }{\delta X_{\mu }}-T^{2}\frac{dX^{\mu }}{d\sigma }%
\frac{dX_{\mu }}{d\sigma }-2iT\left( \chi _{a}\frac{d\chi ^{a}}{d\sigma }+%
\overline{\chi }^{a}\frac{d\overline{\chi }_{a}}{d\sigma }\right) \right]
\label{flatham}
\end{equation}
Substituting the oscillator expansion of equations (\ref{Xosc},\ref{chi},\ref
{chibar}) into equation (\ref{flatham}) gives 
\begin{equation}
\begin{array}{c}
\begin{array}{c}
H=-\left[ \frac{\delta }{\delta X_{0}^{\mu }}\frac{\delta }{\delta X_{0\mu }}%
+\frac{1}{2}\displaystyle\sum\limits_{n>0}n\left( \frac{\delta }{\delta
a_{n}^{\mu }}\frac{\delta }{\delta a_{n\mu }}+\frac{\delta }{\delta 
\widetilde{a}_{n}^{\mu }}\frac{\delta }{\delta \widetilde{a}_{n\mu }}\right)
\right. \qquad \qquad \qquad \\ 
\quad \quad -\frac{1}{2}\displaystyle\sum\limits_{n>0}n\left( a_{n}^{\mu
}a_{n\mu }+\widetilde{a}_{n}^{\mu }\widetilde{a}_{n\mu }\right) -\frac{1}{2}%
\displaystyle\sum\limits_{r\in Z_{0}+\phi }r\left(
c_{r}^{a}d_{r}^{a}-d_{r}^{a}c_{r}^{a}\right)
\end{array}
\\ 
\quad \left. -\frac{i}{2}\displaystyle\sum\limits_{r\in Z_{0}+\phi }r\left( 
\widetilde{c}_{r}^{a}\widetilde{d}_{r}^{a}-\widetilde{d}_{r}^{a}\widetilde{c}%
_{r}^{a}\right) \right] \qquad \hspace{0in}\qquad \qquad \qquad \quad \quad
\end{array}
\label{oscham}
\end{equation}
The momentum $P_{0\mu }$ is identified with $-i\frac{\delta }{\delta
X_{0}^{\mu }}$ and a linear transformation that rotates the fields $\left(
a_{n}^{\mu },\widetilde{a}_{n}^{\mu },\frac{\delta }{\delta a_{n}^{\mu }},%
\frac{\delta }{\delta \widetilde{a}_{n}^{\mu }}\right) $ into $\left( \alpha
_{n}^{\mu },\overline{\alpha }_{n}^{\mu },\alpha _{n}^{\mu \dagger },%
\overline{\alpha }_{n}^{\mu \dagger }\right) $ is performed: 
\begin{equation}
\left( 
\begin{array}{c}
a_{n}^{\mu } \\ 
\widetilde{a}_{n}^{\mu } \\ 
\frac{\delta }{\delta a_{n}^{\mu }} \\ 
\frac{\delta }{\delta \widetilde{a}_{n}^{\mu }}
\end{array}
\right) =\frac{1}{2}\left( 
\begin{array}{cccc}
1 & 1 & 1 & 1 \\ 
-i & i & i & -i \\ 
1 & 1 & -1 & -1 \\ 
-i & i & -i & i
\end{array}
\right) \left( 
\begin{array}{l}
\alpha _{n}^{\mu } \\ 
\overline{\alpha }_{n}^{\mu } \\ 
\alpha _{n}^{\mu \dagger } \\ 
\overline{\alpha }_{n}^{\mu \dagger }
\end{array}
\right)  \label{mattrans}
\end{equation}
It is easy to verify that $\alpha _{n}^{\mu },\alpha _{n}^{\mu \dagger }$and 
$\overline{\alpha }_{n}^{\mu }$ $,\overline{\alpha }_{n}^{\mu \dagger }$
form creation and annihilation operator pairs: 
\begin{eqnarray}
\left[ \alpha _{n}^{\mu },\alpha _{m}^{\nu \dagger }\right] &=&-\delta
_{mn}\delta ^{\mu \nu }  \label{alpha} \\
\left[ \overline{\alpha }_{n}^{\mu },\overline{\alpha }_{n}^{\mu \dagger
}\right] &=&-\delta _{mn}\delta ^{\mu \nu }  \label{alphabar}
\end{eqnarray}
Similarly for the fermions we rotate the fields $\left( c_{r}^{a},\overline{c%
}_{r}^{a},d_{r}^{a},\overline{d}_{r}^{a}\right) $, into $\left( b_{r}^{a},%
\overline{b}_{r}^{a},b_{-r}^{a},\overline{b}_{-r}^{a}\right) $: 
\begin{equation}
\left( 
\begin{array}{l}
c_{r}^{a} \\ 
\overline{c}_{r}^{a} \\ 
d_{r}^{a} \\ 
\overline{d}_{r}^{a}
\end{array}
\right) =\frac{1}{2}\left( 
\begin{array}{cccc}
i & 1 & i & 1 \\ 
-i & 1 & -i & 1 \\ 
-1 & -i & 1 & i \\ 
1 & -i & 1 & i
\end{array}
\right) \left( 
\begin{array}{l}
b_{r}^{a} \\ 
\overline{b}_{r}^{a} \\ 
b_{-r}^{a} \\ 
\overline{b}_{-r}^{a}
\end{array}
\right)  \label{ctrans}
\end{equation}
so that $b_{r}^{a},b_{-r}^{a}$ and $\overline{b}_{r}^{a},\overline{b}%
_{-r}^{a}$ satisfy the (anti)commutation relations: 
\begin{eqnarray}
\left\{ b_{r}^{a},b_{-s}^{b}\right\} &=&\delta ^{ab}\delta _{rs}
\label{bquan} \\
\left\{ \overline{b}_{r}^{a},\overline{b}_{-s}^{b}\right\} &=&\delta
^{ab}\delta _{rs}  \label{bbar}
\end{eqnarray}
In terms of the new operators the Hamiltonian (\ref{oscham}) takes the form 
\begin{eqnarray}
H &=&P_{0\mu }P_{0}^{\mu }+\frac{1}{2}\displaystyle\sum\limits_{n>0}n\left(
\alpha _{n}^{\mu \dagger }\alpha _{n}^{\mu }+\alpha _{n}^{\mu }\alpha
_{n}^{\mu \dagger }+\overline{\alpha }_{n}^{\mu \dagger }\overline{\alpha }%
_{n}^{\mu }+\overline{\alpha }_{n}^{\mu }\overline{\alpha }_{n}^{\mu \dagger
}\right)  \nonumber \\
&&+\frac{1}{2}\displaystyle\sum\limits_{r\in Z_{0}+\phi }r\left(
-b_{r}^{a}b_{-r}^{a}+b_{-r}^{a}b_{r}^{a}-\overline{b}_{r}^{a}\overline{b}%
_{-r}^{a}-\overline{b}_{-r}^{a}\overline{b}_{r}^{a}\right) \qquad \qquad
\label{stringH}
\end{eqnarray}
and the momentum $P$ in (\ref{Peq}) becomes 
\begin{eqnarray}
P &=&\frac{1}{2}\displaystyle\sum\limits_{n>0}n\left( \alpha _{n}^{\mu
\dagger }\alpha _{n}^{\mu }+\alpha _{n}^{\mu }\alpha _{n}^{\mu \dagger }-%
\overline{\alpha }_{n}^{\mu \dagger }\overline{\alpha }_{n}^{\mu }-\overline{%
\alpha }_{n}^{\mu }\overline{\alpha }_{n}^{\mu \dagger }\right)  \nonumber \\
&&\!\!\!\!+\frac{1}{2}\displaystyle\sum\limits_{r\in Z_{0}+\phi }r\left(
-b_{r}^{a}b_{-r}^{a}-b_{-r}^{a}b_{r}^{a}-\overline{b}_{r}^{a}\overline{b}%
_{-r}^{a}+\overline{b}_{-r}^{a}\overline{b}_{r}^{a}\right) \quad \qquad
\qquad \qquad  \label{stringP}
\end{eqnarray}
It can be easily checked that $\frac{1}{2}(H\pm P)$ split into left and
right movers which are functions of $\left( \alpha _{n}^{\mu },\alpha
_{n}^{\mu \dagger },b_{r}^{a},b_{-r}^{a}\right) $ and $\left( \overline{%
\alpha }_{n}^{\mu },\overline{\alpha }_{n}^{\mu \dagger },\overline{b}%
_{r}^{a},\overline{b}_{-r}^{a}\right) $ respectively.

Up to now we have ignored adding the superghost system due to gauge fixing
the two-dimensional metric and gravitino. The gauge fixing terms are given
by \cite{Friedan} 
\begin{equation}
S^{(\mathrm{ghost)}}=-\frac{1}{2\pi }\int d^{2}\xi d\theta _{+}d\theta
_{-}\left( BD_{-}C+\overline{B}D_{+}\overline{C}\right)  \label{ghostaction}
\end{equation}
where the fields $B$ and $C$ and have the component expansions: 
\begin{equation}
\begin{array}{ccc}
B=\beta +i\theta _{+}b & \qquad \overline{B}=\overline{\beta }-i\theta _{-}%
\overline{b} &  \\ 
C=c+i\theta _{+}\gamma & \qquad \overline{C}=\overline{c}-i\theta _{-}%
\overline{\gamma } & 
\end{array}
\end{equation}
satisfying the equations of motion. In component form the ghost action
becomes 
\begin{equation}
S^{(\mathrm{ghost)}}=\frac{1}{2\pi }\displaystyle\int d^{2}\xi (b\partial
_{-}c+\beta \partial _{-}\gamma +\overline{b}\partial _{+}\overline{c}+%
\overline{\beta }\partial _{+}\overline{\gamma })
\end{equation}
The associated ghost-supercurrents are: 
\begin{equation}
\begin{array}{cc}
j_{+}^{(\mathrm{ghost)}} & =-c\partial _{+}\beta +\frac{1}{2}\gamma b-\frac{3%
}{2}\partial _{+}c\beta \\ 
j_{-}^{(\mathrm{ghost)}} & =-\overline{c}\partial _{-}\overline{\beta }+%
\frac{1}{2}\overline{\gamma }\overline{b}-\frac{3}{2}\partial _{-}\overline{c%
}\overline{\beta }
\end{array}
\label{ghostcur}
\end{equation}
The fields $b,c$ and $\beta ,\gamma $ satisfy the quantization conditions: 
\begin{equation}
\begin{array}{cc}
\left\{ \,b(\sigma ,\tau ),c({\sigma }^{\prime },\tau )\right\} \,=2\pi
\delta (\sigma -{\sigma }^{\prime }) &  \\ 
\left[ \beta (\sigma ,\tau ),\gamma ({\sigma }^{\prime },\tau )\right] =2\pi
\delta (\sigma -{\sigma }^{\prime }) & 
\end{array}
\label{bcsystem}
\end{equation}
with similar relations to the conjugate fields. One can define the ghost
Dirac-Ramond operators by 
\begin{equation}
Q_{\pm }^{(\mathrm{ghost)}}=\frac{1}{2}\int\limits_{0}^{2\pi }d\sigma
\,j_{\pm }^{(\mathrm{ghost)}}  \label{ghostcharge}
\end{equation}
which will satisfy 
\begin{equation}
(Q_{\pm }^{(\mathrm{ghost)}})^{2}=H^{(\mathrm{ghost})}\pm P^{(\mathrm{ghost}%
)}  \label{ghostham}
\end{equation}
The Hamiltonian and momenta of the ghost system do not interact with the
rest, but simply add up, allowing for this part to be computed separately,
as it is independent of the background fields: 
\begin{equation}
Q_{\pm }=Q_{\pm }^{(\mathrm{matter)}}+Q_{\pm }^{(\mathrm{ghost)}}
\end{equation}

The system in general consists of states which are eigenvectors of $H\pm P$,
and can be taken as the tensor product of the left-moving sector times the
right-moving sector. The fermionic states are given by the sum of states in
the NS- and R-sectors. Associating the fermionic numbers $F_{L}=F$ and $%
F_{R}=\overline{F}$ with the grading operators, we have $\Gamma =\left(
-1\right) ^{F}$, and $\overline{\Gamma }=\left( -1\right) ^{\overline{F}}$%
which satisfy the commutation relations 
\begin{eqnarray}
\{\Gamma ,Q\} &=&0=[\overline{\Gamma },Q]  \label{gradingp} \\
\{\overline{\Gamma },\overline{Q}\} &=&0=[\Gamma ,\overline{Q}]
\label{gradingm}
\end{eqnarray}
It is possible to repeat the above analysis to determine (perturbatively)
the states of the system when non-trivial backgrounds are present. This will
be important when we put the system in the noncommutative geometry setting.

\section{Noncommutative geometry of the non-linear sigma model}

Most of the considerations of the last section could be looked at from the
non-linear sigma model study and one may ask for the relevance of
noncommutative geometry. The point of view that we like to advance is that
once a spectral triple $({\mathcal{A}},\mathcal{H},D)$ is specified it is
possible to define a noncommutative space and use the tools of
noncommutative geometry \cite{ConnesBook}. A spectral triple $({\mathcal{A}},%
\mathcal{H},D)$ is defined such that $\mathcal{A}$ is an algebra of
operator, $\mathcal{H}$ is the Hilbert space of states on $\mathcal{A}$, and 
$D$ a Dirac operator on $\mathcal{H}$. We shall first briefly review some of
the basic definitions in noncommutative geometry so that we can make the
necessary identifications.

Given a unital involutive algebra $\mathcal{A}$, one can define the
universal space of differential forms 
\begin{equation}
\Omega ^{.}\left( \mathcal{A}\right) =\bigoplus\limits_{n=0}^{\infty }\Omega
^{n}\left( \mathcal{A}\right)
\end{equation}
by setting $\Omega ^{0}\left( \mathcal{A}\right) =\mathcal{A}$ and defining
the linear space $\Omega ^{n}\left( \mathcal{A}\right) $ by 
\begin{equation}
\Omega ^{n}\left( \mathcal{A}\right) =\left\{
\sum\limits_{i}a_{0}^{i}da_{1}^{i}\cdots da_{n}^{i}\,;\,\quad a_{j}^{i}\in 
\mathcal{A},\quad \forall i,j\right\}
\end{equation}
For $\Omega ^{.}\left( \mathcal{A}\right) $ to be a right module, $d$ must
obey Liebniz rule 
\begin{equation}
d\left( a.b\right) =\left( da\right) .b+a.\left( db\right)
\end{equation}
An element of $\Omega ^{n}\left( \mathcal{A}\right) $ is a form of degree $n$%
.

One forms play a special role as components of connections on a line bundle
whose space of action is given by the algebra $\mathcal{A}$. A one-form $%
\rho \in \Omega ^{1}\left( \mathcal{A}\right) $ can be expressed as 
\begin{equation}
\rho =\sum\limits_{i}a^{i}db^{i}
\end{equation}
To analyze the noncommutative space more concretely, we introduce the
notions of a Dirac K-cycle for $\mathcal{A}$ on $\mathcal{H}$. We say that $%
\left( \mathcal{H},D\right) $is a Dirac cycle for $\mathcal{A}$ if there
exists an involutive representation $\pi $ of $\mathcal{A}$ satisfying $\pi
\left( a^{*}\right) =\pi \left( a\right) ^{*}$ with the properties that $\pi
\left( a\right) $ and $\left[ D,\pi \left( a\right) \right] $ are bounded
operators on $\mathcal{H}$ for all $a\in \mathcal{A}$, and that $\left(
D^{2}+1\right) ^{-1}$ is a compact operator on $\mathcal{H}$. A K-cycle $%
\left( \mathcal{H},D\right) $ is even if there exists a unitary involution $%
\Gamma $ on $\mathcal{H}$ with $\Gamma ^{*}=\Gamma ^{-1}=\Gamma $ such that 
\begin{equation}
\left[ \Gamma ,\pi \left( a\right) \right] =0,\quad \forall a\in \mathcal{A}
\end{equation}
\begin{equation}
\left\{ \Gamma ,D\right\} =0
\end{equation}
A representation $\pi $ of $\Omega ^{.}\left( \mathcal{A}\right) $ on $%
\mathcal{H}$ is defined by 
\begin{equation}
\pi \left( \sum\limits_{i}a_{0}^{i}da_{1}^{i}\cdots da_{n}^{i}\right)
=\sum\limits_{i}\pi \left( a_{0}^{i}\right) \left[ D,\pi \left(
a_{1}^{i}\right) \right] \cdots \left[ D,\pi \left( a_{n}^{i}\right) \right]
\end{equation}
In addition, one can define vector bundles $E,$ connections $\nabla $ on $E$%
, the associated curvature $R=-\nabla ^{2}$, etc. Many of the tools
available for Riemannian geometry could be generalized to the noncommutative
case. For details see \cite{CFFgravity} \cite{CF}.

In the case of a supersymmetric system in two-dimensions there are as many
operators $Q$ as there are supersymmetries. The corresponding noncommutative
spaces would have more structure and could be classified according to the
degree of supersymmetry. The simplest possibility is to have $(1,0)$
supersymmetry as there will be only one charge $Q$ which could be identified
with $D$. In the case of $(1,1)$ supersymmetry we have seen that we have the
relations 
\begin{equation}
Q_{\pm }^{2}=H\pm P
\end{equation}
therefore, provided that the states are restricted to those with $P=0$, one
can make the identifications $Q_{1}=d,Q_{2}=d^{*}$ where $Q_{\pm }=Q_{1}\pm
Q_{2}.$

For the example studied in the last section we have $\mathcal{A}=C^{\infty
}(\Omega (M))$, the algebra of continuous functions on the loop space over
the manifold $M$ \cite{W85}. Elements of the algebra are functionals of the
form $f\,[X^{\mu }(\sigma )]$ where $\sigma $ parametrizes the circle. The
operators $Q_{\pm }$ are used to define $H$ and $P$. The Hilbert space is
the sum of the NS- and R-sectors, each of which can be further divided into
bosonic and fermionic states such that the operators $Q_{\pm }|\Omega
\rangle $ would be fermionic if $|\Omega \rangle $ is bosonic. One may ask
whether it is possible to avoid the use of noncommutative geometry and work
within the context of loop space. There are many difficulties in working
with infinite dimensional loop spaces, in contrast to noncommutative
geometry where a rich structure is available. There is also an advantage in
treating this model with the noncommutative geometric tools as this would
allow us to consider, in the future, more complicated examples which could
only be treated by noncommutative geometric methods.

To illustrate, consider the operator 
\begin{equation}
D=Q_{+}+Q_{-}  \label{Dirac1}
\end{equation}
which satisfies $D^{2}=H+P$. From the property 
\begin{equation}
e^{i\epsilon \,P}f\,\,[X^{\mu }(\sigma )]e^{-i\epsilon \,P}=f\,\,[X^{\mu
}(\sigma +\epsilon )]
\end{equation}
it is clear that $P$ generates diffeomorphisms on the circle. Restricting to
states which are reparametrization invariance, the operator $D^{2}$ acting
on this subspace will be equal to $H$. Then it is possible to build the
universal space of differential forms. A one-form is given by 
\begin{eqnarray}
\pi (\rho ) &=&\sum_{i}f^{\,\,i}\left[ D,g^{i}\right]  \nonumber \\
&=&\sum_{i}\int d\sigma \,\left[ f\,^{i}[X]\left( \psi _{+}^{\mu }+\psi
_{-}^{\mu }\right) \frac{\delta g^{i}}{\delta X^{\mu }}\right] _{P=0}
\end{eqnarray}
The main difficulty for $\left( 1,1\right) $ supersymmetry is the
availability of two Dirac operators, the first is in (\ref{Dirac1}) and the
second is $\overline{D}=i\left( Q_{+}-Q_{-}\right) $ satisfying $\overline{D}%
^{2}=H-P$. We shall restrict ourselves to the case where only $D$ is used,
and geometric objects are evaluated on the subspace where $P=0$ so that the
two operators become equivalent. We now proceed to apply these ideas.

\section{The superstring spectral action}

In a recent proposal \cite{ACAC} it was suggested that the action must be
spectral. One starts with the spectral data and then identify the symmetry
transformations as automorphisms of the algebra $\mathcal{A}$. If the Dirac
operator is of some special type, one can by allowing transformations on
elements of the algebra to generate more general operators corresponding to
generic cases. This is, however, a complicated process, and is usually
difficult. As we saw, for the Dirac operators of the $(1,1)$ superstring we
were only able to find the correct form by deriving the supercurrent of the
system. It will be important to generalize this further by allowing for all
possible excitations, massless or massive to enter. In this paper we shall
be mainly concerned with studying the dynamics of the massless states.

The spectral action must be of the form 
\begin{equation}
I=\mathrm{Tr}_{NS\oplus R}F(Q_{+},Q_{-},\Gamma ,\overline{\Gamma })|_{P=0}
\label{spectaction}
\end{equation}
restricted to states with $P=0$. It is a challenge to find the appropriate
function $F$ and this is the question we will now face. In the case of the
noncommutative space of the standard model \cite{group} it turned out that
in the low-energy domain (lower than the planck scale) the exact form of the
function $F$ is not important except for the requirement that the function
is well behaved. By expanding the function in terms of its Mellin transform,
the problem reduces to the heat kernel expansion of the pseudo-differential
operator which contains all the geometric invariants. Therefore in this case
the exact form of the function is only important to determine the coupling
constants in front of the geometric invariants. If one wants to study the
dynamics at very high energies the heat kernel expansion breaks down and the
exact form of the function becomes essential. We shall see that even if the
function $F$ is known, evaluating the action could still be a formidable job.

Usually one starts with the expression \cite{ACAC} 
\[
\mathrm{Tr}F(D^{2})=\displaystyle\sum\limits_{s>0}F_{s}\mathrm{Tr}(D^{2}) 
\]
then by using the definition of the Mellin transform one can show that 
\begin{equation}
\mathrm{Tr}F(D^{2})=\displaystyle\sum\limits_{n>0}F_{n}a_{n}(D^{2})
\label{heatkernel}
\end{equation}
\begin{eqnarray}
F_{0} &=&\int\limits_{0}^{\infty }uF(u)du \\
F_{2} &=&\int\limits_{0}^{\infty }F(u)du \\
F_{2(n+1)} &=&F^{(n)}(0)
\end{eqnarray}
The coefficients $a_{n}$ are defined by \cite{Gilkey} 
\begin{equation}
\mathrm{Tr}(e^{-tD^{2}})=\displaystyle\sum\limits_{n>0}t^{\frac{n-m}{d}%
}a_{n}(D^{2})  \label{expan}
\end{equation}
which is the heat-kernel expansion of the operator $D$. In the case under
consideration we need to evaluate expressions of the form 
\begin{equation}
\mathrm{Tr}(e^{-\theta _{2}H})=\mathrm{Tr}(e^{-\theta
_{2}(Q_{+}Q_{-}+Q_{-}Q_{+})})|_{P=0}  \label{partition}
\end{equation}
restricted to states with $P=0$. This can be achieved by inserting a delta
function 
\begin{equation}
\int d\theta _{1}e^{-i\theta _{1}(Q_{+}^{2}+Q_{+}^{2})}  \label{delta}
\end{equation}
Inserting identity (\ref{delta}) in the trace formula (\ref{partition}) we
get 
\begin{equation}
\mathrm{Tr}(e^{-i\theta Q^{2}+i\overline{\theta }\overline{Q}^{2}})
\end{equation}
where $\theta =\theta _{1}+i\theta _{2}$, with the appropriate integration
over $\theta $.

To find the correct form of the function $F$ we notice that the one-loop
vacuum amplitude of the superstring in a flat background can be calculated
exactly. The Hamiltonian and momentum in this case simplify to equations (%
\ref{stringH}, \ref{stringP}). The path integral expression of the one-loop
amplitude, is related to the partition function , in the case when the
two-dimensional surface is a torus \cite{Luis83}. The result is modular
invariant, and therefore consistent (free of anomalies) if the dimension of
the target space is $D=10$. We also have to set $T=\frac{1}{4\pi l_{s}^{2}}$
where $l_{s}$ is the string length scale (which will be set to one). To
project the non-physical states out (or equivalently, require modular
invariance when the two-dimensional surface has genus greater than or equal
to two) one must have the partition function \cite{SW} 
\begin{equation}
I=\int \frac{d\theta d\overline{\theta }}{\theta _{2}^{2}}\mathrm{Tr}\left|
\sum_{NS\oplus R}\left( e^{2\pi i(\theta Q^{2}}(-1)^{\epsilon }(1-\Gamma
)\right) \right| ^{2}  \label{postulate}
\end{equation}
where $\theta $ is the modular parameter $\epsilon =0$ for the NS sector,
and $\epsilon =1$ for the Ramond sector over the states taken in the trace.
This partition function, has space-time supersymmetry as can be verified by
counting the number of fermionic and bosonic states (massive as well as
massless) and showing they are the same \cite{GSW}. The total partition
function, including the ghosts is 
\begin{equation}
\int \frac{d\theta d\overline{\theta }}{\theta _{2}^{2}}\int \frac{d^{8}p}{%
(2\pi )^{8}}e^{-2\pi p_{2}^{2}\theta _{2}}\frac{1}{\left| \eta \left( \theta
\right) \right| ^{24}}\left| \left( \mathcal{\vartheta }_{3}^{4}(0|\theta )-%
\mathcal{\vartheta }_{4}^{4}(0|\theta )^{4}-\mathcal{\vartheta }%
_{2}^{4}(0|\theta )^{4}-\mathcal{\vartheta }_{1}^{4}(0|\theta )\right)
\right| ^{2}  \label{paf}
\end{equation}
As expected, because of supersymmetry, the partition function vanishes as
follows from the theta functions identity. The $dp$ integration gives a
factor of $\frac{1}{\theta _{2}^{4}}$ which renders the partition function (%
\ref{paf}) modular invariant.

The ghost contributions cancel the contributions of two bosonic and two
fermionic coordinates. Since the superghost part is independent of the
background, these contributions would be the same even in a curved
background.

It is tempting to postulate that $I$ in (\ref{postulate}) is the spectral
action of the superstring when $Q$ and $\overline{Q}$ include the
fluctuations of a general background geometry. The remaining part of the
paper is to give supporting evidence for this conjecture.

Since we have only derived the expressions for $Q$ and $\overline{Q}$ when
the two-dimensional surface is a sphere, one must work out the necessary
changes when the sphere is replaced with a torus. The two-dimensional metric
is \cite{pilch} 
\begin{equation}
g_{\alpha \beta }=\left( 
\begin{array}{ll}
\theta _{1}^{2}+\theta _{2}^{2} & \theta _{1} \\ 
\theta _{1} & 1
\end{array}
\right)
\end{equation}
Denoting 
\begin{eqnarray*}
z &=&\sigma +\theta \tau ,\quad \overline{z}=\sigma +\overline{\theta }\tau ,
\\
\partial &=&2i\theta _{2}\partial _{z},\quad \overline{\partial }=-2i\theta
_{2}\partial _{\overline{z}},
\end{eqnarray*}
and after rescaling of $\psi ^{a\pm }\rightarrow \frac{1}{\sqrt{\theta _{2}}}%
\psi ^{a\pm }$ one finds out that the non-linear sigma model $S$ takes the
same form as in (\ref{surface}) with the replacements $T\rightarrow \frac{1}{%
\theta _{2}^{2}}T$ , $\partial _{+}\rightarrow \partial $ and $\partial
_{-}\rightarrow \overline{\partial }$. First one derives the currents from
this action, then after quantization, the charges would take exactly the
same form as in equations (\ref{jplus},\ref{jminus}) if one further rescale $%
X^{\mu }\rightarrow \frac{1}{\sqrt{\theta _{2}}}X^{\mu }$ and $T\rightarrow
\theta _{2}T$. The charges are then found to be independent of $\theta $.
The $X^{\mu }\,$rescaling can be understood from the requirement that the
supersymmetry transformations keep the same form as in equation (\ref{trans}%
). Therefore, we can make the equivalent statement that by rescaling the
superfields by $\Phi ^{\mu }\rightarrow \frac{1}{\sqrt{\theta _{2}}}\Phi
^{\mu }$ and the torsion by $T\rightarrow \frac{1}{\theta _{2}}T$ the
charges become independent of $\theta $. This shows that the Hamiltonian and
momentum are independent of $\theta $. At this point we can postulate that
the spectral action (\ref{postulate}) should be modular invariant. Having
demonstrated that the form of $H\pm P$ is independent of the modular
parameter it follows that all the dependence on $\theta $ is included in the
coefficient in the partition function. The challenging problem now is to
evaluate the proposed action (\ref{postulate}) and show that it is the
correct one. To get a handle on this problem we shall adopt the background
field method to expand the fields in normal coordinates. The points $%
X_{0}^{\mu }$ and $X_{0}^{\mu }+\Pi ^{\mu }$ are connected through a
geodesic with the tangent to the geodesic at $X_{0}^{\mu }$ denoted by $\xi
^{\mu }$. The expansion for the metric is \cite{Luis81} 
\begin{equation}
G_{\mu \nu }\left[ X_{0}+\Pi \right] =G_{\mu \nu }\left[ X_{0}\right] -\frac{%
1}{3}R_{\mu \rho \nu \sigma }\left[ X_{0}\right] \xi ^{\rho }\xi ^{\sigma }-%
\frac{1}{3!}D_{\kappa }R_{\mu \rho \upsilon \sigma }\left[ X_{0}\right] \xi
^{\kappa }\xi ^{\rho }\xi ^{\sigma }+\cdots  \label{back}
\end{equation}
where the vector $\xi ^{\mu }$ contain only the oscillator parts. Using this
form of the expansion it is possible to order the terms in $H$ and $P$
according to their degree in $\xi ^{\mu }$. Similarly we can expand the
other fields such as $B_{\mu \nu }$ in normal coordinates 
\begin{equation}
B_{\mu \nu }\left[ X_{0}+\Pi \right] =B_{\mu \nu }\left[ X_{0}\right]
+D_{\rho }B_{\mu \nu }\left[ X_{0}\right] \xi ^{\rho }+\cdots
\end{equation}
The fermionic fields $\chi ^{a}$ and $\overline{\chi }^{a}$ are separated
into zero modes and non-zero modes in the R-sector. The zero order part of
the Hamiltonian in the Ramond sector is (setting $T=\frac{1}{4\pi }$) : 
\begin{eqnarray}
(H\pm P)_{\mathrm{R}}^{0} &=&-\left[ \nabla _{0}^{a}\nabla _{0a}+\omega
_{0b}^{\ \ ab}\nabla _{0a}+4\chi _{0}^{a}\bar{\chi}_{0}^{b}\chi _{0}^{c}\bar{%
\chi}_{0}^{d}R_{abcd}^{0}\right.  \nonumber \\
&&+\frac{2}{3}(\chi _{0}^{a}\bar{\chi}_{0}^{b}\bar{\chi}_{0}^{c}\bar{\chi}%
_{0}^{d}+\bar{\chi}_{0}^{a}\chi _{0}^{b}\chi _{0}^{c}\chi _{0}^{d}+\bar{\chi}%
_{0}^{a}\chi _{0}^{b}\chi _{0}^{c}\bar{\chi}_{0}^{d})\nabla _{0a}H_{0bcd} 
\nonumber \\
&&+(\chi _{0}^{b}\chi _{0}^{c}+\bar{\chi}_{0}^{b}\bar{\chi}%
_{0}^{c})H_{0bca}\nabla _{0a}  \nonumber \\
&&\left. +\frac{1}{3}(\chi _{0}^{a}\chi _{0}^{b}\bar{\chi}_{0}^{c}\bar{\chi}%
_{0}^{d}+\chi _{0}^{a}\bar{\chi}_{0}^{b}\bar{\chi}_{0}^{c}\chi _{0}^{d}+\chi
_{0}^{a}\bar{\chi}_{0}^{b}\bar{\chi}_{0}^{c}\bar{\chi}_{0}^{d}+\bar{\chi}%
_{0}^{a}\bar{\chi}_{0}^{b}\chi _{0}^{c}\chi _{0}^{d})H_{0ab}^{\ \,\,\,\
e}H_{0ecd}\right] \qquad  \label{Hramond}
\end{eqnarray}
where $\nabla _{0a}=e_{a}^{\mu }\left[ X_{0}\right] \left( \frac{\delta }{%
\delta X_{0}^{\mu }}+\omega _{\mu ab}[X_{0}]\left( \chi _{0}^{a}(\sigma )%
\overline{\chi }_{0}^{b}(\sigma )+\overline{\chi }_{0}^{a}(\sigma )\chi
_{0}^{b}(\sigma )\right) \right) $ is the connection on the spin-manifold $M$%
. We can use the identity 
\begin{equation}
\nabla _{0}^{a}\nabla _{0a}+\omega _{0b}^{\ \ ab}\nabla _{0a}=G_{0}^{\mu \nu
}\left[ X_{0}\right] \nabla _{0\mu }\nabla _{0\nu }-\Gamma _{0}^{\mu }\nabla
_{0\mu }
\end{equation}
to change the operator into the conventional type. Here $\Gamma _{0}^{\mu }$
is a contraction of the Christoffel symbol. In the NS-sector there are no
zero modes for the fermions $\chi ^{a}$ and the zero-order Hamiltonian
reduces to 
\begin{equation}
\left( H\pm P\right) _{\mathrm{NS}}^{0}=-\left( G^{\mu \nu }[X_{0}]\partial
_{0\mu }\partial _{0\nu }-\Gamma _{0}^{\mu }\partial _{0\mu }\right)
\end{equation}
Needless to say, the higher order terms are more complicated, and increase
dramatically in number. The most important pieces for our purposes are the
quadratic terms in the fluctuations $\xi ^{\mu }$ and $\chi ^{a}-\chi
_{0}^{a}$. This enables us to define the ground state perturbatively and
evaluate the trace.

We first group the terms which are quadratic in $\xi ^{\mu }$ and $\frac{%
\partial }{\partial \xi ^{\mu }}$ then expand these in terms of $\sigma $,
then perform the $\sigma $ integration to obtain the quadratic bosonic part
valid for both the R-sector and the NS-sector: 
\begin{equation}
\frac{1}{2}G_{\mu \nu }[X_{0}]\displaystyle\sum\limits_{n>0}n\left( \alpha
_{n}^{\mu \dagger }\alpha _{n}^{\nu }+\alpha _{n}^{\mu }\alpha _{n}^{\nu
\dagger }+\overline{\alpha }_{n}^{\mu \dagger }\overline{\alpha }_{n}^{\nu }+%
\overline{\alpha }_{n}^{\mu }\overline{\alpha }_{n}^{\nu \dagger }\right)
\end{equation}
where $\alpha _{n}^{\mu }$ and $\alpha _{n}^{\nu \dagger }$ are defined as
in equation (\ref{mattrans}) except that $\frac{\delta }{\delta a_{n}^{\mu }}
$ is replaced by $G^{\mu \nu }[X_{0}]\frac{\delta }{\delta a_{n}^{\nu }}$ to
make the expression covariant. The treatment of the pure quadratic fermionic
part does not change because the spinors $\chi ^{a}$ have a tangent space
index. The contribution to the Hamiltonian is given as in (\ref{stringH}): 
\begin{equation}
\frac{1}{2}\displaystyle\sum\limits_{r\in Z_{0}+\phi }r\left(
-b_{r}^{a}b_{-r}^{a}+b_{-r}^{a}b_{r}^{a}-\overline{b}_{r}^{a}\overline{b}%
_{-r}^{a}-\overline{b}_{-r}^{a}\overline{b}_{r}^{a}\right) \qquad
\end{equation}
where $\phi =0$ in the R-sector and $\phi =\frac{1}{2}$ in the NS-sector. In
evaluating the trace it is necessary to put the operators in the form $%
\alpha _{n}^{\mu }\alpha _{n}^{\nu \dagger }$ and $b_{-r}^{a}b_{r}^{a}$ and
this gives rise to infinities and need normal ordering. We adopt the Riemann
zeta-function regularization and write 
\begin{equation}
\zeta (s)=\sum\limits_{n=1}^{\infty }n^{-s}
\end{equation}
which has a unique analytic continuation at $s=-1$ where it has the value $%
\zeta (-1)=-\frac{1}{12}$. The contributions from the bosons and fermions in
the R-sector is $(1+\frac{1}{2})\frac{D}{2}$ $\zeta (-1)$ while from the
ghosts and superghosts is $(-13+\frac{11}{2})\zeta (-1).$ This shows that
the ground state energy in the R-sector is zero for $D=10.$ Similar
considerations show that the vacuum energy of the ground state in the
NS-sector is $6\zeta (-1)$ for $D=10$ \cite{GSW}.

By expanding the metric $G_{\mu \nu }$ as in equation (\ref{back}) then
writing $\xi ^{\mu }$ and $\frac{\partial }{\partial \xi ^{\mu }}$ in terms
of creation and annihilation operators, we notice that the expression $%
\int\nolimits_{0}^{2\pi }d\sigma \left( G^{\mu \nu }\nabla _{\mu }\nabla
_{\nu }\right) $ contains the term 
\begin{equation}
\sum\limits_{n>0}\frac{1}{n}(\alpha _{n}^{\rho }+\alpha _{n}^{\rho })(\alpha
_{n}^{\kappa }+\alpha _{n}^{\kappa })R_{\quad \,\,\,\,\rho \kappa }^{0\mu
\nu }\nabla _{0\mu }\nabla _{0\nu }
\end{equation}
which gives a logarithmic contribution to the trace of the form 
\begin{equation}
\frac{2}{3\epsilon }R^{0\mu \nu }\nabla _{0\mu }\nabla _{0\nu }
\end{equation}
and this term could be absorbed by a redefinition of the metric (to lowest
orders) 
\begin{equation}
G_{0}^{\mu \nu }\rightarrow G_{0}^{\mu \nu }+\frac{2}{3\epsilon }R^{0\mu \nu
}
\end{equation}
We therefore find the phenomena of the renormalization of the fields in the
non-linear sigma models \cite{Luis81}. There are many more terms of
quadratic order. The main difficulty arises because of the terms quartic in
the fermions. In the case when the fermions are truncated to one chirality
the quartic terms drop out and it becomes possible to calculate the various
traces. The different pieces were calculated by Schellekens and Warner \cite
{SCW} (with the $B_{\mu \nu }$ field set to zero) and used by Witten \cite
{W87} in the computation of the elliptic genus.

In the conjectured spectral action (\ref{postulate}) we shall meet the
following four types of terms. These are in the NS-NS, NS-R, R-NS and R-R
which will be denoted respectively by $(--),(-+),(+-),(++).$ First we give
the fermionic contributions in the various sectors. In the $\left( --\right) 
$ sector we have, in the approximation where the quartic terms are ignored, 
\begin{eqnarray}
P_{--}^{(f)} &=&\mathrm{Tr}(e^{i\theta (H+P)_{\mathrm{NS}}^{\prime }}) 
\nonumber \\
&=&\prod\limits_{\beta =1}^{5}\left( \frac{\mathcal{\vartheta }_{3}\left( 
\frac{x_{\beta }}{2\pi i}\mid \theta \right) }{\eta (\theta )}\right) 
\label{Pmm}
\end{eqnarray}
where the $x_{\beta }$ are the eigenvalues of the curvature two-form
modified by the contributions of the torsion pieces. In $(H+P)^{\prime }$
defined in equation (\ref{Pmm}) the pure bosonic part and the zero-order
terms are excluded, and where $\mathcal{\vartheta }_{i}$ $(i=1,\ldots ,4)$
are the generalized Jacobi theta functions. Similarly, and in the same
approximation, we have in the NS-R sector: 
\begin{eqnarray}
P_{-+}^{(f)} &=&\mathrm{Tr}(e^{i\theta (H+P)_{\mathrm{NS}}^{\prime
}}(-1)^{F})  \nonumber \\
&=&\prod\limits_{\beta =1}^{5}\left( \frac{\mathcal{\vartheta }_{4}\left( 
\frac{x_{\beta }}{2\pi i}\mid \theta \right) }{\eta (\theta )}\right) 
\end{eqnarray}
In the R-NS sector we have 
\begin{eqnarray}
P_{+-}^{(f)} &=&\mathrm{Tr}(e^{i\theta (H+P)_{\mathrm{R}}^{\prime }}) 
\nonumber \\
&=&\prod\limits_{\beta =1}^{5}\left( \frac{\mathcal{\vartheta }_{2}\left( 
\frac{x_{\beta }}{2\pi i}\mid \theta \right) }{\eta (\theta )}\right) 
\end{eqnarray}
and finally, in the R-R sector we get: 
\begin{eqnarray}
P_{++}^{(f)} &=&\mathrm{Tr}(e^{i\theta (H+P)_{\mathrm{R}}^{\prime }}(-1)^{F})
\nonumber \\
&=&\prod\limits_{\beta =1}^{5}\left( \frac{\mathcal{\vartheta }_{1}\left( 
\frac{x_{\beta }}{2\pi i}\mid \theta \right) }{\eta (\theta )}\right) 
\end{eqnarray}
The contributions of the bosonic terms are: 
\begin{eqnarray}
P^{(b)} &=&\mathrm{Tr}(e^{i\theta (H+P)_{\mathrm{b}}^{\prime }})  \nonumber
\\
&=&\prod\limits_{\beta =1}^{5}\left( \frac{{2\sinh (}\frac{1}{2}x_{\beta
})\eta (\theta )}{\mathcal{\vartheta }_{1}\left( \frac{x_{\beta }}{2\pi i}%
\mid \theta \right) }\right) 
\end{eqnarray}
where $(H+P)_{\mathrm{b}}^{\prime }$ is the bosonic part of $(H+P)$ without
the zero-order terms. This partition function has to be multiplied by the
Dirac genus $\widehat{A}\left( R,H\right) $ which is given by 
\begin{equation}
\widehat{A}=\prod\limits_{\beta =1}^{5}\frac{\frac{1}{2}x_{\beta }}{\sinh (%
\frac{1}{2}x_{\beta })}
\end{equation}
Now it must be pointed out that when higher order terms are taken into
account, the partition function will not be given by theta functions and all
the above expressions have to be calculated perturbatively. The lowest order
corrections come from the traces due to the zero-order terms: $\mathrm{Tr}%
(e^{-\theta _{2}H_{\mathrm{NS}}^{0}})\mathrm{\ }$and $\mathrm{Tr}(e^{-\theta
_{2}H_{\mathrm{R}}^{0}})$ .These are of the form $\mathrm{Tr}(e^{-\tau _{2}%
\mathcal{P}})$ which could be calculated using the heat kernel expansion: 
\cite{Gilkey} 
\begin{equation}
\mathrm{Tr}(e^{-\theta _{2}\mathcal{P}})=\sum_{n=0}^{\infty }a_{n}(\mathcal{P%
})\,\theta _{2}^{\frac{n-D}{2}}
\end{equation}
where $a_{n}(\mathcal{P})$ are the Seeley-de Wit coefficients corresponding
to the operator $\mathcal{P}$ and $D=10$ is the dimension of the target
manifold. Writing the operator $\mathcal{P}$ in the form 
\begin{equation}
P=-\left( G^{\mu \nu }\partial _{\mu }\partial _{\nu }.\mathcal{I}+\mathcal{A%
}^{\mu }\partial _{\mu }+\mathcal{B}\right)   \label{elliptic}
\end{equation}
where $\mathcal{A}^{\mu }$ , $\mathcal{B}^{\mu }$ and $\mathcal{I}$ are
matrices of the same dimension, $\mathcal{I}$ being the identity matrix. We
shall need only the first two coefficients $a_{0}(X,P)$ and $a_{2}(X,P)$
which are given by 
\begin{equation}
a_{0}(X,P)=\frac{1}{\left( 4\pi \right) ^{\frac{D}{2}}}\mathrm{Tr}(\mathcal{I%
})
\end{equation}
\begin{equation}
a_{2}(X,P)=\frac{1}{\left( 4\pi \right) ^{\frac{D}{2}}}\mathrm{Tr}(\frac{R}{6%
}\mathcal{I}+\mathcal{E})
\end{equation}
where $\mathcal{E}$ is related to $\mathcal{A}^{\mu }$ and $\mathcal{B}$ by 
\begin{equation}
\mathcal{E=B}-G^{\mu \nu }(\partial _{\mu }\omega _{\nu }^{\prime }+\omega
_{\mu }^{\prime }\omega _{\nu }^{\prime }-\Gamma _{\mu \nu }^{\rho }\omega
_{\rho }^{\prime })  \label{Eequation}
\end{equation}
\begin{equation}
\omega _{\mu }^{\prime }=\frac{1}{2}G_{\mu \nu }(\mathcal{A}^{\nu }+\Gamma
^{\nu }\mathcal{I})  \label{omegap}
\end{equation}
Applying these formulas to the operator $H_{\mathrm{NS}}^{0}$ it is easy to
see that 
\begin{equation}
a_{2}(X,P)=\frac{1}{\left( 4\pi \right) ^{\frac{D}{2}}}\mathrm{Tr}(\frac{R}{6%
}\mathcal{I})
\end{equation}
because in this case $\mathcal{A}^{\nu }=-\Gamma ^{\nu }\mathcal{I}$ and $%
\omega _{\mu }^{\prime }=0$ which implies that $\mathcal{E}=0.$ Calculating $%
\mathrm{Tr}(e^{-\theta _{2}H_{\mathrm{R}}^{0}})$ is more involved. First by
comparing equation (\ref{Hramond}) with equation (\ref{elliptic}) we find
that 
\begin{equation}
\mathcal{A}^{\mu }=2\omega _{0}^{\mu }-\Gamma ^{\mu }+(\chi _{0}^{b}\chi
_{0}^{c}+\bar{\chi}_{0}^{b}\bar{\chi}_{0}^{c})H_{0bc}^{\quad \mu }
\end{equation}
\begin{eqnarray}
\mathcal{B} &=&\left( \partial ^{\mu }\omega _{0\mu }+\omega _{0}^{\mu
}\omega _{0\mu }-\Gamma _{0}^{\mu }\omega _{0\mu }-4\chi _{0}^{a}\bar{\chi}%
_{0}^{b}\chi _{0}^{c}\bar{\chi}_{0}^{d}R_{abcd}^{0}\right.   \nonumber \\
&&+(\chi _{0}^{\nu }\chi _{0}^{\rho }+\bar{\chi}_{0}^{\nu }\bar{\chi}%
_{0}^{\rho })H_{0\nu \rho }^{\mu }\omega _{0\mu }  \nonumber \\
&&+\frac{2}{3}(\chi _{0}^{a}\bar{\chi}_{0}^{b}\bar{\chi}_{0}^{c}\bar{\chi}%
_{0}^{d}+\bar{\chi}_{0}^{a}\chi _{0}^{b}\chi _{0}^{c}\chi _{0}^{d}+\bar{\chi}%
_{0}^{a}\chi _{0}^{b}\chi _{0}^{c}\bar{\chi}_{0}^{d})\nabla _{0a}H_{0bcd} \\
&&\left. +\frac{1}{3}(\chi _{0}^{a}\chi _{0}^{b}\bar{\chi}_{0}^{c}\bar{\chi}%
_{0}^{d}+\chi _{0}^{a}\bar{\chi}_{0}^{b}\bar{\chi}_{0}^{c}\chi _{0}^{d}+\chi
_{0}^{a}\bar{\chi}_{0}^{b}\bar{\chi}_{0}^{c}\bar{\chi}_{0}^{d}+\bar{\chi}%
_{0}^{a}\bar{\chi}_{0}^{b}\chi _{0}^{c}\chi _{0}^{d})H_{0ab}^{\ \
\,\,\,e}H_{0ecd}\right)   \nonumber
\end{eqnarray}
Using equations (\ref{Eequation},\ref{omegap}) we can evaluate 
\begin{equation}
\omega _{\mu }^{\prime }=\omega _{0\mu }+\frac{1}{2}(\chi _{0}^{\nu }\chi
_{0}^{\rho }+\bar{\chi}_{0}^{\nu }\bar{\chi}_{0}^{\rho })H_{0\mu \nu \rho }
\end{equation}
\begin{eqnarray}
\mathcal{E} &=&\left[ -4\chi _{0}^{a}\bar{\chi}_{0}^{b}\chi _{0}^{c}\bar{\chi%
}_{0}^{d}R_{abcd}^{0}+\frac{1}{2}(\chi _{0}^{\nu }\chi _{0}^{\rho }+\bar{\chi%
}_{0}^{\nu }\bar{\chi}_{0}^{\rho })D^{\mu }H_{0\mu \nu \rho }\right.  
\nonumber \\
&&+\left( \frac{1}{3}(\chi _{0}^{a}\chi _{0}^{b}\bar{\chi}_{0}^{c}\bar{\chi}%
_{0}^{d}+\chi _{0}^{a}\bar{\chi}_{0}^{b}\bar{\chi}_{0}^{c}\chi _{0}^{d}+\chi
_{0}^{a}\bar{\chi}_{0}^{b}\bar{\chi}_{0}^{c}\bar{\chi}_{0}^{d}+\bar{\chi}%
_{0}^{a}\bar{\chi}_{0}^{b}\chi _{0}^{c}\chi _{0}^{d})\right.   \nonumber \\
&&-\left. \left. \left( \chi _{0}^{a}\chi _{0}^{b}\chi _{0}^{c}\chi _{0}^{d}+%
\overline{\chi }_{0}^{a}\bar{\chi}_{0}^{b}\bar{\chi}_{0}^{c}\overline{\chi }%
_{0}^{d}\right) \right) H_{0ab}^{\ \ \,\,\,\,e}H_{0ecd}\right] 
\label{EEquation}
\end{eqnarray}
From the definition of the traces $\mathrm{Tr}(\chi _{0}^{a}\bar{\chi}%
_{0}^{b})=\frac{1}{4}\delta ^{ab}$ ,one can show that 
\begin{equation}
\mathrm{Tr(}\chi _{0}^{a}\chi _{0}^{b}\bar{\chi}_{0}^{c}\bar{\chi}_{0}^{d})=-%
\frac{1}{16}\left( \delta _{c}^{a}\delta _{d}^{b}-\delta _{d}^{a}\delta
_{c}^{b}\right) \mathrm{Tr}(\mathcal{I})  \label{trace}
\end{equation}
Using equations (\ref{trace},\ref{EEquation}) we deduce that 
\begin{equation}
\mathrm{Tr}(\mathcal{E})=-\frac{1}{4}(R_{0}+\frac{1}{6}H_{0\mu \nu \rho
}H_{0}^{\mu \nu \rho })\mathrm{Tr}(\mathcal{I})
\end{equation}
and this in turn gives $a_{2}(X,H_{\mathrm{R}}^{0}):$%
\begin{equation}
a_{2}(X,H_{\mathrm{R}}^{0})=\frac{1}{({2\pi })^{5}}(-\frac{1}{12}R_{0}-\frac{%
1}{24}H_{0\mu \nu \rho }H_{0}^{\,\mu \nu \rho })  \label{atwo}
\end{equation}
It is clear that $a_{0}(X,H_{\mathrm{R}}^{0})=a_{0}(X,H_{\mathrm{NS}}^{0}).$
The cross term of the NS and R-sectors gets contributions in the heat-kernel
expansion from the $H_{\mathrm{NS}}^{0}$ operator only.

The last step is to list the contributions of the ghost and superghost
sectors and these are well known. For the $b-c$ system the partition
function is the same in the NS and R-sectors. It is given by 
\begin{equation}
P^{b-c}=\mathrm{Tr}(e^{i\theta (H+P)^{b-c}})=\left( \eta (\theta )\right)
^{2}
\end{equation}
The contributions of the $\beta -\gamma $ system depends on the boundary
conditions of the NS and R-sectors. These are 
\begin{equation}
P_{--}^{\beta -\gamma }=\mathrm{Tr}(e^{i\theta (H+P)_{\mathrm{NS}}^{\beta
-\gamma }})=\left( \frac{\eta (\theta )}{\mathcal{\vartheta }_{3}(0\mid
\theta )}\right) 
\end{equation}
\begin{equation}
P_{-+}^{\beta -\gamma }=\mathrm{Tr}(e^{i\theta (H+P)_{\mathrm{NS}}^{\beta
-\gamma }}(-1)^{F})=\left( \frac{\eta (\theta )}{\mathcal{\vartheta }%
_{4}(0\mid \theta )}\right) 
\end{equation}
\begin{equation}
P_{+-}^{\beta -\gamma }=\mathrm{Tr}(e^{i\theta (H+P)_{\mathrm{R}}^{\beta
-\gamma }})=\left( \frac{\eta (\theta )}{\mathcal{\vartheta }_{2}(0\mid
\theta )}\right) 
\end{equation}
\begin{equation}
P_{++}^{\beta -\gamma }=\mathrm{Tr}(e^{i\theta (H+P)_{\mathrm{R}}^{\beta
-\gamma }}(-1)^{F})=\lim_{\nu \rightarrow 0}\left( \frac{\eta (\theta )}{%
\mathcal{\vartheta }_{1}(\nu \mid \theta )}\right) 
\end{equation}
the limit in the last expression is given by $(2\nu )^{2}\eta (\theta )^{4}$
(one should not be worried about the zero mode of the $\beta -\gamma $
system in the R-sector as these will be cancelled by the lowest order terms
of two of the fermions). We are now ready to write down the spectral action,
corresponding to the one-loop string effective action, where not only the
zero-modes are taken into account but also the oscillators: 
\[
I=\int\limits_{M}d^{10}X_{0}\sqrt{G\left[ X_{0}\right] }\left| \left(
\prod\limits_{\beta =1}^{5}\left( \frac{{x}_{\beta }\eta (\theta )}{\mathcal{%
\vartheta }_{1}\left( \frac{x_{\beta }}{2\pi i}\mid \theta \right) }\right)
\right) \eta ^{2}(\theta )\right| ^{2}\theta _{2}.
\]
\begin{equation}
\left( \left( \frac{a_{0}^{(\mathrm{NS})}}{\theta _{2}^{5}}+\frac{a_{2}^{(%
\mathrm{NS})}}{\theta _{2}^{4}}+\cdots \right) \left( \left| A\right| ^{2}-(A%
\overline{B}+\overline{A}B)\right) +\left( \frac{a_{0}^{(\mathrm{R})}}{%
\theta _{2}^{5}}+\frac{a_{2}^{(\mathrm{R})}}{\theta _{2}^{4}}+\cdots \right)
\left| B\right| ^{2}\right) 
\end{equation}
where we have denoted 
\begin{equation}
A=\left( \left( \frac{\eta (\theta )}{\mathcal{\vartheta }_{3}(0\mid \theta )%
}\right) ^{2}F_{3}-\left( \frac{\eta (\theta )}{\mathcal{\vartheta }%
_{4}(0\mid \theta )}\right) ^{2}F_{4}\right) 
\end{equation}
\begin{equation}
B=\left( \left( \frac{\eta (\theta )}{\mathcal{\vartheta }_{2}(0\mid \theta )%
}\right) ^{2}F_{2}-\left( \frac{\eta (\theta )}{\mathcal{\vartheta }%
_{1}(0\mid \theta )}\right) ^{2}F_{1}\right) 
\end{equation}
and the functions $F_{i},\,i=1,\ldots ,4,$ are given by 
\begin{equation}
F_{i}=\prod\limits_{\beta =1}^{5}\left( \frac{\mathcal{\vartheta }_{i}\left( 
\frac{x_{\beta }}{2\pi i}\mid \theta \right) }{\eta (\theta )}\right)
+\cdots 
\end{equation}
It is not difficult to check that in the limit when the quartic fermionic
terms are terminated the action reduces to 
\begin{equation}
I=\int\limits_{M}d^{10}X_{0}\sqrt{G\left[ X_{0}\right] }\int \frac{d\theta d%
\overline{\theta }}{\theta _{2}^{2}}\left| \left( \prod\limits_{\beta
=1}^{5}\left( \frac{{x}_{\beta }\eta (\theta )}{\mathcal{\vartheta }%
_{1}\left( \frac{x_{\beta }}{2\pi i}\mid \theta \right) }\right) \right)
\eta ^{2}(\theta )\right| ^{2}\frac{1}{\theta _{2}^{4}}\left| A-B\right| ^{2}
\end{equation}
which is modular invariant as can be seen from the transformations under the
modular group 
\begin{equation}
\mathcal{\vartheta }_{i}\left( \frac{\nu }{c\theta +d}\mid \frac{a\theta +b}{%
c\theta +d}\right) =\varepsilon \exp \left( \frac{i\pi c\nu ^{2}}{c\theta +d}%
\right) \mathcal{\vartheta }_{j}(\nu \mid \theta )
\end{equation}
where $\varepsilon $ is a complicated phase as function of $i$ and $j$. This
form of the action contains terms which are at least of order $R^{2}$ and $%
H^{4}$ and misses an important piece due to ignoring the quartic terms as
these could give rise to the $R$ and $H^{2}$ terms as seen from equation (%
\ref{atwo}). The perturbative form of this action is obtained by using the
following expansions of the Jacobi-theta functions \cite{Scho} 
\begin{equation}
\left( \frac{\mathcal{\vartheta }_{1}\left( \nu \mid \theta \right) }{\nu 
\mathcal{\vartheta }_{1}^{^{\prime }}\left( 0\mid \theta \right) }\right)
=\exp \left( -\sum\limits_{k=1}^{\infty }\frac{1}{2k}\nu
^{2k}G_{2k}(q^{2})\right) 
\end{equation}
\begin{equation}
\left( \frac{\mathcal{\vartheta }_{2}\left( \nu \mid \theta \right) }{%
\mathcal{\vartheta }_{2}\left( 0\mid \theta \right) }\right) =\exp \left(
\sum\limits_{k=1}^{\infty }\frac{1}{2k}\nu ^{2k}\left(
G_{2k}(q^{2})-2^{2k}G_{2k}(q^{4})\right) \right) 
\end{equation}
\begin{equation}
\left( \frac{\mathcal{\vartheta }_{3}\left( \nu \mid \theta \right) }{%
\mathcal{\vartheta }_{3}\left( 0\mid \theta \right) }\right) =\exp \left(
\sum\limits_{k=1}^{\infty }\frac{1}{2k}\nu ^{2k}\left(
G_{2k}(q^{2})-G_{2k}(-q)\right) \right) 
\end{equation}
\begin{equation}
\left( \frac{\mathcal{\vartheta }_{3}\left( \nu \mid \theta \right) }{%
\mathcal{\vartheta }_{3}\left( 0\mid \theta \right) }\right) =\exp \left(
\sum\limits_{k=1}^{\infty }\frac{1}{2k}\nu ^{2k}\left(
G_{2k}(q^{2})-G_{2k}(q)\right) \right) 
\end{equation}
where $G_{2k}(q)$ are the Eisenstein functions and $q=e^{i\pi \theta }.$
This gives the lowest order term 
\begin{equation}
\int\limits_{M}d^{10}X_{0}\sqrt{G\left[ X_{0}\right] }\int \frac{d\theta d%
\overline{\theta }}{\theta _{2}^{2}}\frac{1}{\theta _{2}^{4}}\frac{1}{\left|
\eta \left( \theta \right) \right| ^{24}}\left( \mathcal{\vartheta }%
_{3}^{4}(0|\theta )-\mathcal{\vartheta }_{4}^{4}(0|\theta )-\mathcal{%
\vartheta }_{2}^{4}(0|\theta )-\mathcal{\vartheta }_{1}^{4}(0|\theta
)\right) 
\end{equation}
and this is zero as would be expected from for the vanishing of the
cosmological constant in superstring theory. The next low-order term is 
\begin{eqnarray}
&&\int\limits_{M}d^{10}X_{0}\sqrt{G\left[ X_{0}\right] }\int \frac{d\theta d%
\overline{\theta }}{\theta _{2}^{2}}\frac{1}{\theta _{2}^{3}}\left( \left( 
\frac{R}{6}\right) \left( \left| L_{1}\right| ^{2}-\left( L_{1}\overline{%
L_{2}}+\overline{L_{1}}L_{2}\right) \right) \right.   \nonumber \\
&&\qquad \qquad \qquad \qquad \qquad \qquad \qquad \left. +\left( \frac{R}{6}%
-\frac{R}{4}-\frac{H_{\mu \nu \rho }^{2}}{24}\right) \left| L_{2}\right|
^{2}\right)   \label{sugra}
\end{eqnarray}
where 
\begin{equation}
L_{1}=\mathcal{\vartheta }_{3}^{4}(0|\theta )-\mathcal{\vartheta }%
_{4}^{4}(0|\theta ),\quad L=\mathcal{\vartheta }_{2}^{4}(0|\theta )-\mathcal{%
\vartheta }_{1}^{4}(0|\theta )
\end{equation}
After the integration over the modular parameter is performed the action (%
\ref{sugra}) becomes proportional to the action

\begin{equation}
I\varpropto -\int d^{10}X_{0}\sqrt{G[X_{0}]}(\frac{1}{4}R[X_{0}]+\frac{1}{24}%
H_{0\mu \nu \rho }H_{0}^{\mu \nu \rho })+\cdots
\end{equation}
and this is known to agree, including all coefficients, with the
supergravity action in ten-dimensions of a metric and antisymmetric tensor.%
\cite{acdten} \cite{west}

It is of course not possible to calculate the spectral action (\ref
{postulate}) in closed form using a perturbative expansion only. What one
hopes is to use the requirement of modular invariance of the spectral action
and the limit we found here to search for the appropriate form. In other
words we expect the full spectral action to be of the form 
\begin{equation}
I=\int\limits_{M}d^{10}X_{0}\sqrt{G\left[ X_{0}\right] }\int \frac{d\theta d%
\overline{\theta }}{\theta _{2}^{2}}\left| \left( \prod\limits_{\beta =1}^{5}%
\frac{{x}_{\beta }\eta (\theta )}{\mathcal{\vartheta }_{1}\left( \frac{%
x_{\beta }}{2\pi i}\mid \theta \right) }\right) \eta ^{2}(\theta )\right|
^{2}\frac{1}{\theta _{2}^{4}}\Phi \left( \theta ,\overline{\theta }%
,R,H\right)
\end{equation}
where $\Phi $ is a function such that under the modular transformations $%
\theta \rightarrow \frac{a\theta +b}{c\theta +d}$, $x_{\beta }\rightarrow 
\frac{x_{\beta }}{c\theta +b}$ it is invariant up to a phase 
\begin{equation}
\Phi \rightarrow \left| \exp \left( -\frac{i\pi c\sum\limits_{\beta
=1}^{5}\left( \frac{x_{\beta }}{2\pi i}\right) ^{2}}{c\theta +d}\right)
\right| ^{2}\Phi
\end{equation}
It will take more work to calculate exactly the next leading order, to help
single out uniquely the function $\Phi .$

\section{Conclusions}

\smallskip At present there are many attempts to investigate the structure
of space-time at very high energies. A common feature to all these ideas is
that at planckian energies there will be uncertainty in defining space-time
points, and a richer physical and mathematical structure is needed. One of
the leading candidates for such a description is string theory which made a
considerable progress in the last ten years. What is still lacking in string
theory is a geometrical picture, and the accompanying geometric tools
needed. On the other hand, there was also considerable progress in the last
few years in noncommutative geometry as formulated by Alain Connes \cite
{ConnesBook}. In noncommutative geometry one finds the tools that allows to
handle geometric spaces which could not be studied otherwise. Even the
simplest noncommutative space that one considers, consisting of a product of
a manifold times a discrete set of points and taking a simple matrix
algebra, reproduces the standard model of particle physics with all its
details.

In a recent work by Alain Connes and the author \cite{ACAC},it was
conjectured that information about the geometry and its dynamics are
contained in the spectral triple. Besides the geometric space that one can
define using the spectral triple it is possible, using the \emph{spectral
action principle,} to determine precisely the dynamical structure of the
theory. The symmetry associated with the spectral action is the \emph{%
automorphism of the algebra.} This bold conjecture was tested on the well
established noncommutative geometry of the standard model, and it was shown
that the present particle physics spectrum of quarks and leptons fixes
uniquely the geometry and its dynamics, and an arbitrary spectral action
reproduces, in its lowest order terms, the standard model Lagrangian.

In this paper, we have applied the same ideas to the noncommutative space
constructed from the non-linear supersymmetric sigma model. What makes
supersymmetry essential is that the conserved supersymmetric charges could
be identified with Dirac operators. The algebra is the algebra of continuous
functions on the loop space over the target space manifold, and the Hilbert
space is the Hilbert space of states, massless and massive. The problem we
tackled in this paper is the derivation of the supercharges for a non-linear
sigma model in an arbitrary curved background with torsion and the
determination of the Hilbert space of states which could only be determined
perturbatively. After quantization, the conserved charges become Dirac
operators. The spectral action is then the trace of a function of the Dirac
operators which depends functionally on the background fields. These could
be expanded in Fourier modes, and one has to sum over an infinite set of
oscillators. Therefore, the problem here is unlike the case of
point-particle operators where one can evaluate the action using a heat
kernel expansion, even if one does not know the function. For operators on
loop space, one encounters infinite sums and it is difficult to evaluate the
action even if one knows the spectral function. Fortunately, one knows that
the spectral action should describe the dynamics of the background fields,
which in the case at hand are the metric and antisymmetric tensor. One also
knows that in the critical dimension $D=10$, the cosmological constant
vanishes. The physical space must be restricted to zero momentum states so
as to enforce reparametrization invariance. This lead us to the conjecture
that the spectral action is given by the partition function of the one-loop
vacuum amplitude of the superstring. This idea is then tested, and it is
found that when the terms with the quartic fermion interactions are ignored,
the various pieces involved in the partition function are known. But
ignoring the quartic fermionic terms would miss the lowest order terms in
the effective action. The lowest order terms could be computed by evaluating
the heat-kernel expansion of the operators taking the fermionic zero modes
into account, and oscillator modes should be included. Putting all this
information together, we evaluated the spectral action to lowest order in
the perturbative expansion. It is shown that this reproduces the correct
limit in a highly non-trivial way. What remains to be done is to determine
the exact (non-perturbative) form of the partition function guided by the
limits we found and by the requirement that the spectral function is modular
invariant. This appears to be a difficult task, but one hopes that by
computing the next leading order of the spectral action in perturbation
theory it might be possible to determine this function uniquely.

It is also a difficult exercise to compute the spectral action in an
arbitrary background including the dilaton, and the space-time
supersymmetric vacuum so that a space-time gravitino background, as well as
a two and three forms would be included. The difficulty is because
space-time supersymmetry can only be made explicit by invoking the
Green-Schwarz superstring and $\kappa $ symmetry \cite{GSW} \cite{Lerche},
but world-sheet supersymmetry would not be manifest. These points are not
attempted here but deserve further study.

\end{document}